\documentclass[a4paper,12pt]{article}

\usepackage[dvips]{graphicx}

\title{
Elliptic Flow Based on a Relativistic\\
Hydrodynamic Model
}
\author{
Tetsufumi Hirano\thanks{E-mail: hirano@hep.phys.waseda.ac.jp}
}

\begin{document}
\maketitle

\textit{Department of Physics, Waseda University, Tokyo 169-8555, Japan}
\vspace{36pt}

\abstract
{Based on the (3+1)-dimensional hydrodynamic model, the space-time evolution
of hot and dense nuclear matter produced in non-central relativistic
heavy-ion collisions is discussed.
The elliptic flow parameter $v_2$ is obtained by Fourier analysis of the
azimuthal distribution of pions and protons which are emitted
from the freeze-out hypersurface. 
As a function of rapidity, the pion and proton elliptic flow
parameters both have a peak at midrapidity.
}

\newpage
One of the main goals in relativistic heavy-ion physics is the creation of
a quark-gluon plasma (QGP) and the determination of its equation of state
(EoS) \cite{QM97}.
It is therefore very important to study collective flow in
non-central collisions, such as directed or elliptic flow \cite{OLLIQM97}.
Recently experimental data concerning collective flow in semi-central
collisions at SPS energies has been reported
\cite{NA49,WA98,CERES}. This data should be analysed using
various models.
Some groups \cite{LIU,HEISEL,SOFF} have used their microscopic
transport models to analyse the collective flow obtained by the
NA49 Collaboration \cite{NA49}. 
In this paper we investigate collective flow, especially elliptic flow, in
terms of a relativistic hydrodynamic model.

In non-central collisions elliptic flow arises due to the fact that the
spatial overlap region of two colliding nuclei in the transverse plane has
an ``almond shape". That is, 
the hydrodynamical flow becomes larger along the short axis than 
along the
long axis because the pressure gradient is larger in that direction.
Therefore this \textit{spatial} anisotropy causes the nuclear
matter to also have \textit{momentum} anisotropy.
Consequently, the azimuthal distribution may carry information about the
pressure of the nuclear matter produced in the early stage of
the heavy-ion collisions \cite{OLLI}.

The relativistic hydrodynamical equations for a perfect fluid represent
energy-momentum conservation
\begin{eqnarray}
\partial_\mu T^{\mu \nu} & = & 0,\\
T^{\mu \nu} & = & (E+P)u^\mu u^\nu-Pg^{\mu \nu}
\end{eqnarray}
and baryon density conservation
\begin{eqnarray}
\partial_\mu n_{\mathrm{\small{B}}}^\mu & = & 0,\\
n_{\mathrm{\small{B}}}^\mu & = & n_{\mathrm{\small{B}}} u^\mu,
\end{eqnarray}
where $E$, $P$, $n_{\mathrm{\small{B}}}$ and $u^\mu$ are, respectively,
the energy density, pressure, baryon density and local four velocity.
We numerically solve these equations without assuming cylindrical symmetry
\cite{RISCHKE,NONAKA} by specifying the model EoS and we obtain the space-time dependent thermodynamical variables and the four velocity.

We use the following models of the EoS with a phase transition.
 Hagedorn's statistical bootstrap model \cite{HAGE} with Hagedorn
temperature $T_{\mathrm{\small{H}}}=155$ MeV is employed for the
hadronic phase. 
We directly use the integral representation of the solution of the bootstrap
equation \cite{HAGE2} instead of using the very famous hadronic mass spectrum,
$\exp(m/T_\mathrm{\small{H}})$, which is the \textit{asymptotic} solution of
this equation.
It is well known that this model has a limited temperature
range, i.e., the energy
density and pressure diverge at $T_{\mathrm{\small{H}}}$.
This singularity, however, disappears when an exclude volume
approximation \cite{KAPU} (with a Bag constant $B^{\frac{1}{4}}=230$ MeV)
is associated with the Hagedorn model.
In the QGP phase, we use massless free u, d and s-quarks and the 
gluon gas model for
simplicity.
The two equations of state are matched by imposing Gibbs' condition for phase
equilibrium.
Consequently we obtain a first order phase transition model which has a
critical temperature $T_{\mathrm{\small{C}}} = 159$ MeV and a
mixed phase pressure of $P_{\mathrm{\small{mix}}} = 70.9$
MeV/fm$^3$ at zero baryon density.

We mention our numerical algorithm for the relativistic hydrodynamic model.
It is known that the Piecewise Parabolic Method (PPM) \cite{PPM} is very robust
scheme for the non-relativistic gas equation
 with a shock wave. We have extended
 the PPM scheme of Eulerian hydrodynamics to the relativistic hydrodynamical equation.
Note that this is a higher order extension of the piecewise
\textit{linear} method \cite{rHLLE}.

Assuming non-central Pb+Pb collisions at SPS energy, we choose very simple
formulas for the initial condition at the initial (or passage) time
$t_0=2r_0/(\gamma v) \sim 1.4$ fm ($r_0$, $\gamma$ and $v$ are,
respectively, the nuclear radius, Lorentz factor and  the velocity of
a spectator in the center of mass system)
\begin{eqnarray}
E(x,y,z) & = & E_1(z)
\theta(\tilde{z}_0-z)\theta(z+\tilde{z}_0) \rho(r_\mathrm{p})
\rho(r_\mathrm{t}), \\
n_{\mathrm{\small{B}}}(x,y,z) & = &
n_{\mathrm{\small{B1}}}(z) \theta(\tilde{z}_0-z)\theta(z+\tilde{z}_0)
\rho(r_\mathrm{p}) \rho(r_\mathrm{t}),\\
v_{\mathrm{\small{z}}}(x,y,z) & = & v_0
\tanh(z/z_0)\nonumber \\
& \times &
\theta(\tilde{z}_0-z)\theta(z+\tilde{z}_0)\rho(r_\mathrm{p})
\rho(r_\mathrm{t}),
\end{eqnarray}
\noindent
where $\theta(z)$ is the step function, $\rho(r)$ is the Woods-Saxon
parameterization in the transverse direction,
\begin{eqnarray}
\rho(r) & = &
\frac{1}{\exp\left(\frac{r-r_0}{\delta_{\mathrm{\small{r}}}} \right)+1},
\end{eqnarray}
\noindent
$E_1(z)$ is Bjorken's solution \cite{BJOR} and the $z$ dependence of
the baryon density $n_{\mathrm{\small{B1}}}(z)$ is taken from Ref.~\cite{SOLL}
\begin{eqnarray}
E_1(z) \!& = &\! E_0 \times
\left(\frac{\sqrt{t_0^2-z^2}}{t_0}\right)^{-\frac{4}{3}},\\
n_{\mathrm{\small{B1}}}(z) \!& = &\! \kappa \times0.17
\frac{\sqrt{t_0^2-z^2}}{t_0}.
\end{eqnarray}
\begin{figure}[htbp]
\begin{center}
\includegraphics[width=12cm]{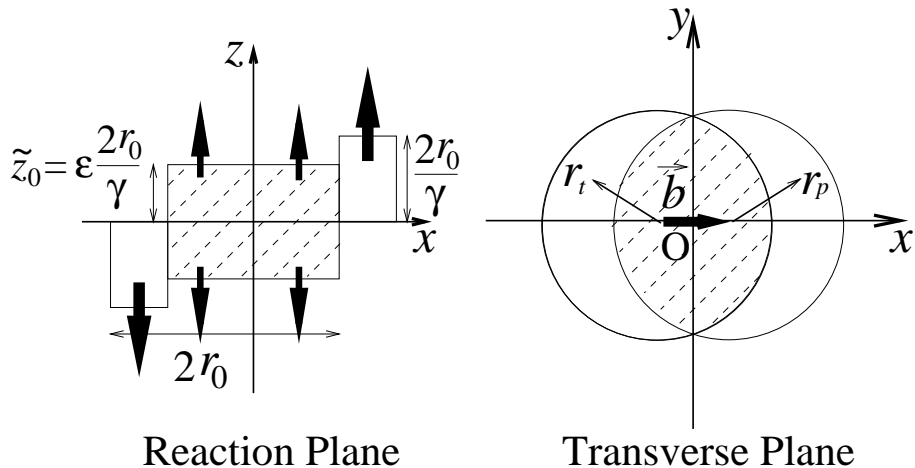}
\caption{Schematic view of the initial geometry in the center of mass system.
The left figure shows the reaction plane and the right the transverse plane.
The initial condition is in the region with slanting lines. $\vec{b}$ is
the impact parameter vector. $r_\mathrm{p}$ and $r_\mathrm{t}$
are respectively the
distances from the center of the projectile and the target
nucleus in the transverse plane.}
\label{fig:initial}
\end{center}
\end{figure}
See also Fig.~\ref{fig:initial}.
We have employed Bjorken's longitudinal solution just as an initial
condition. This is in contrast to Ref.~\cite{OLLI,SHURYAK}, in which 
Bjorken's boost-invariant solution was used as an assumption and the
hydrodynamical equation was numerically solved only in the transverse plane.

At relativistic energies the Lorentz-contracted spectators leave the
interaction region after $\sim 1$ fm, we therefore assume the hydrodynamical
description is valid only in the overlap region and neglect the interaction
between the spectators and the fluid. Therefore we can say that our model gives
a good description only in the vicinity of the midrapidity region and fails to
reproduce \textit{directed} flow at present.
It may be possible to treat this problem if we use a hadronic cascade model
for both spectators and particles emitted from the freeze-out hypersurface,
together with the hydrodynamic model.

There are four initial (and adjustable) parameters in our hydrodynamic
model: 1) the energy density at $z=0$, $E_0=2500$ MeV/fm$^3$, 2) the factor
in the baryon density distribution $\kappa = 2.5$, 3) the initial
longitudinal factor $\varepsilon=0.9$ and 4) the ``diffuseness parameter"
$\delta_{\mathrm{\small{r}}} = 0.3$ fm.
In the present analysis we select these values `by hand', i.e., we
guess them.
\begin{figure}[htbp]
\includegraphics[height=8cm]{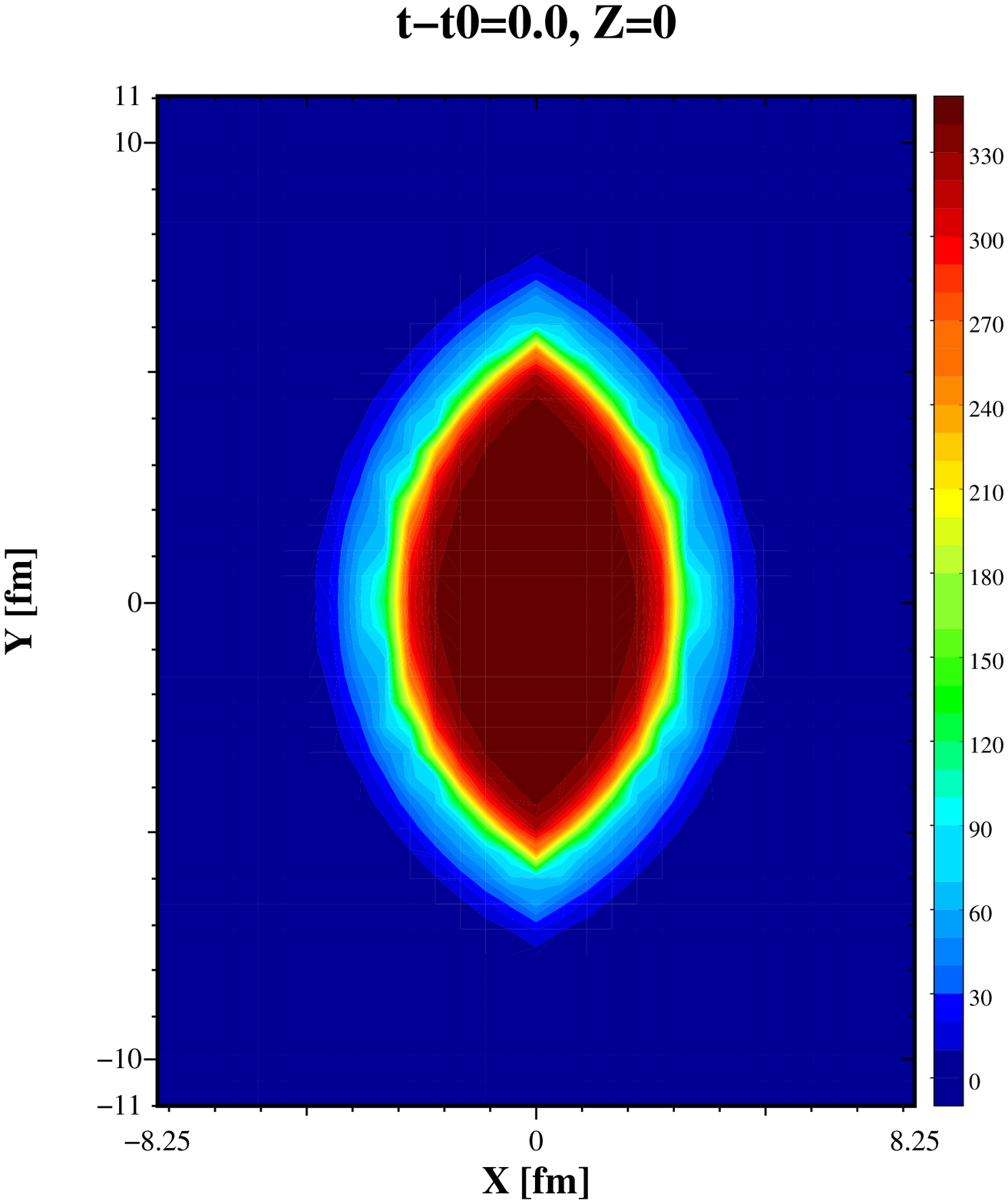}
\hspace{10pt}
\includegraphics[height=8cm]{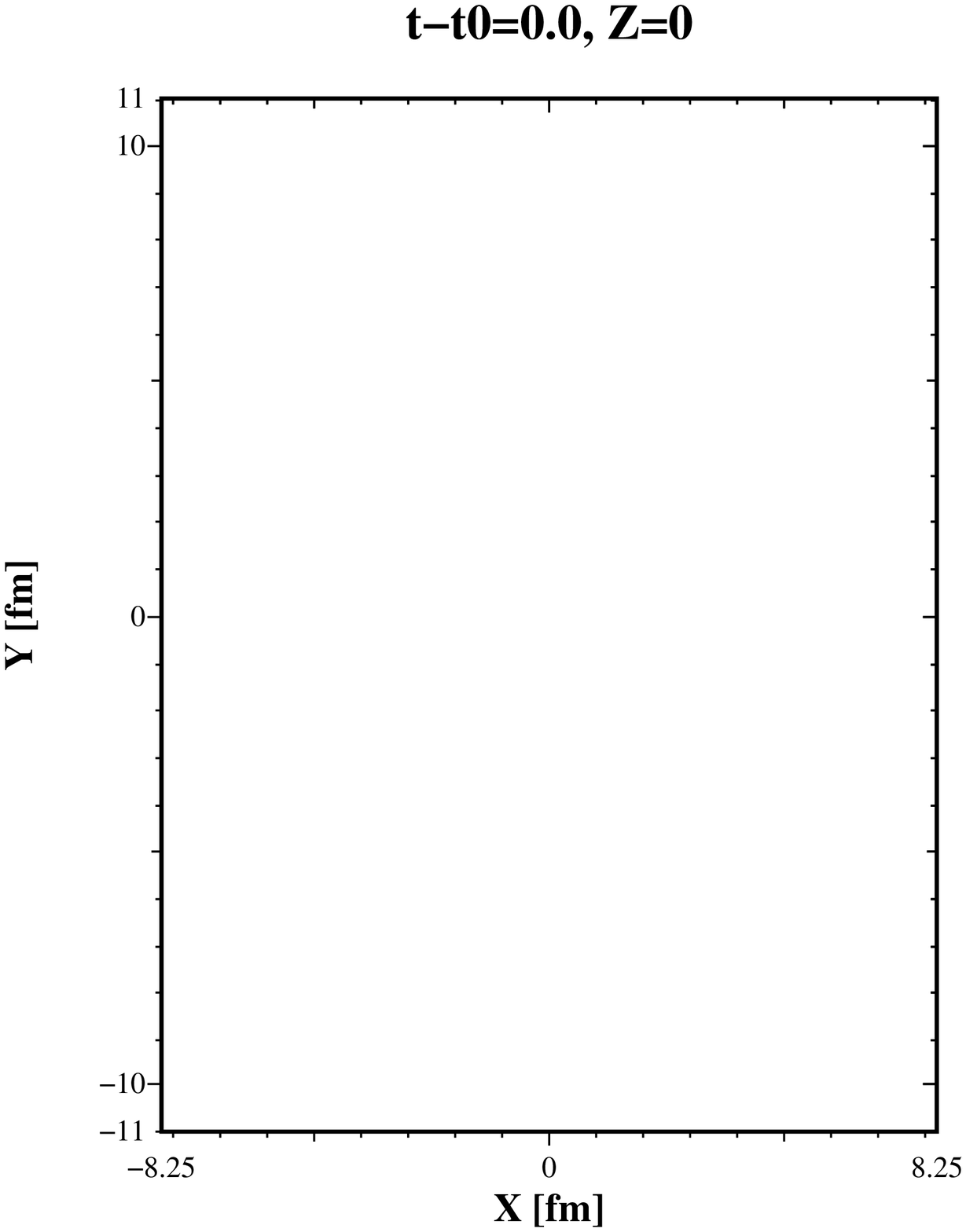}

\vspace{20pt}
\includegraphics[height=8cm]{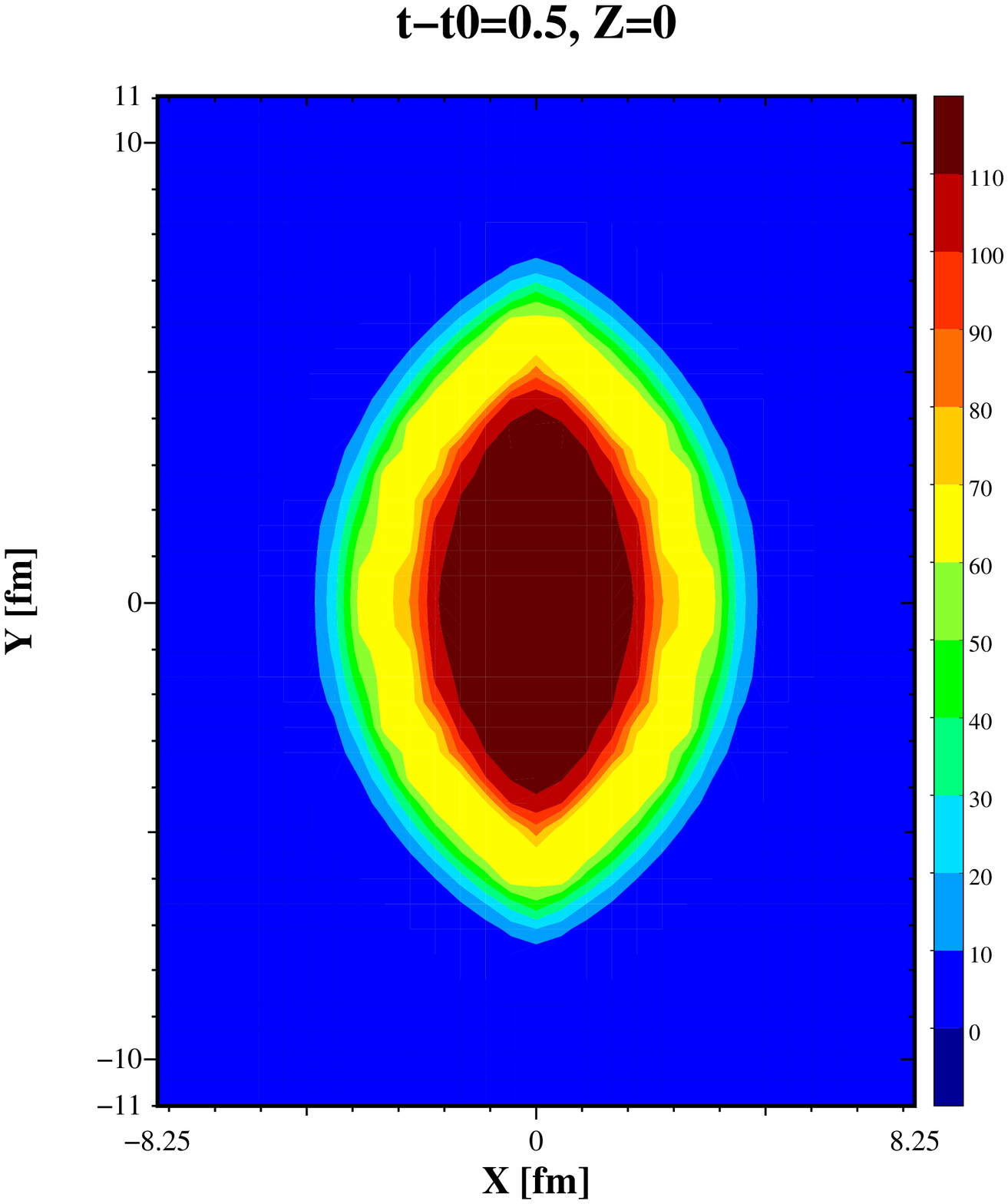}
\hspace{10pt}
\includegraphics[height=8cm]{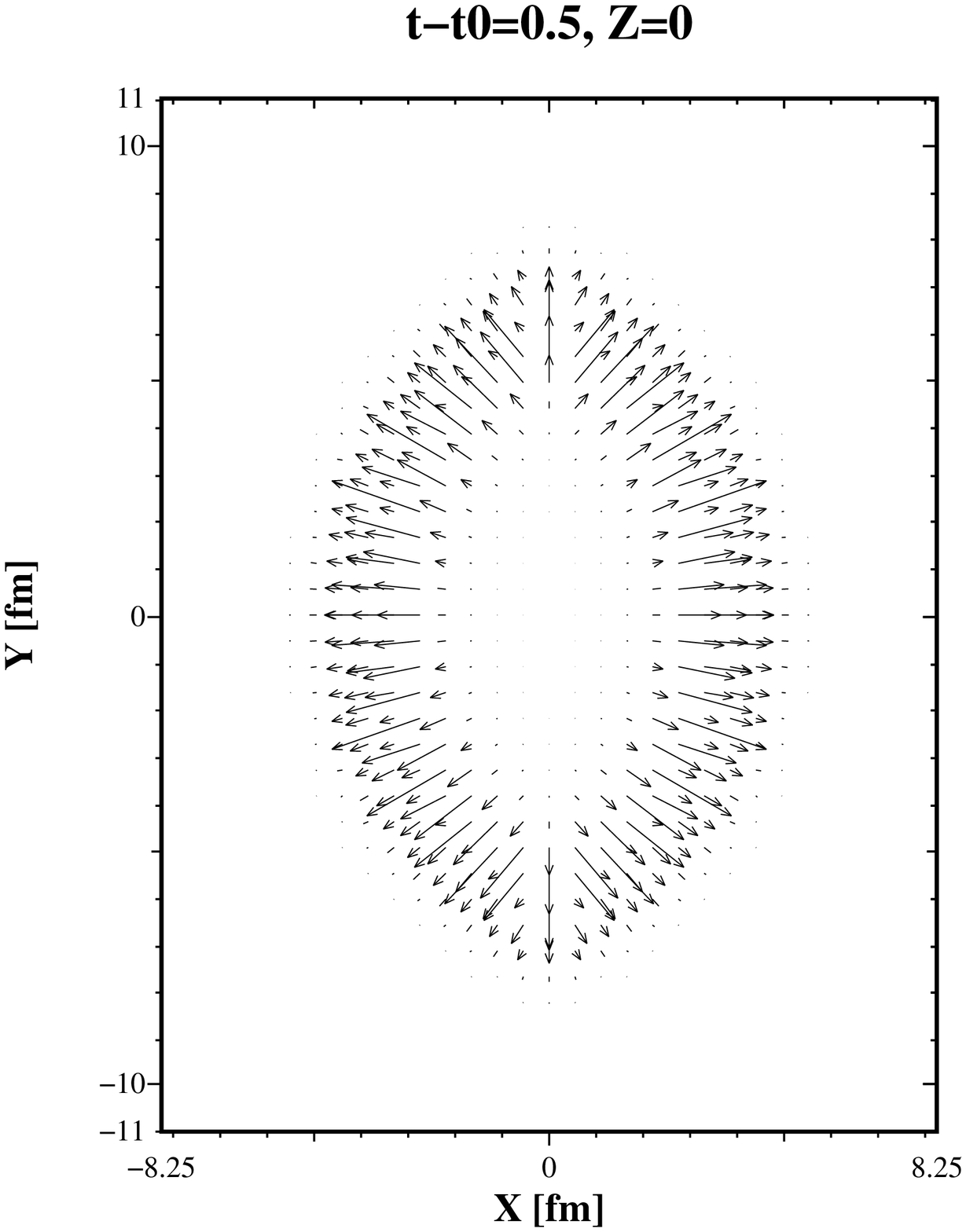}
\begin{center}
(To be continued on next page.)
\end{center}
\end{figure}
\begin{figure}[htbp]
\includegraphics[height=8cm]{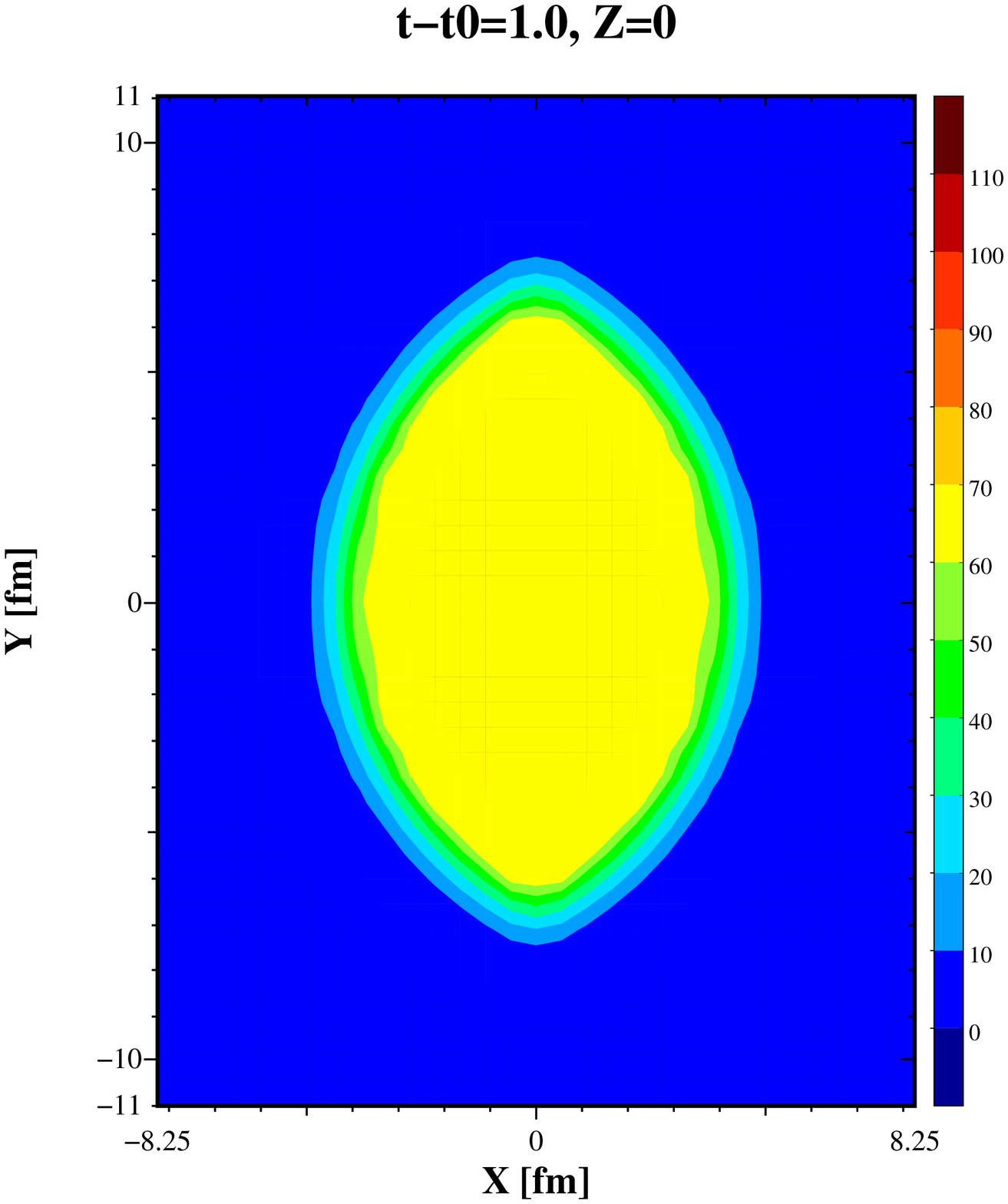}
\hspace{10pt}
\includegraphics[height=8cm]{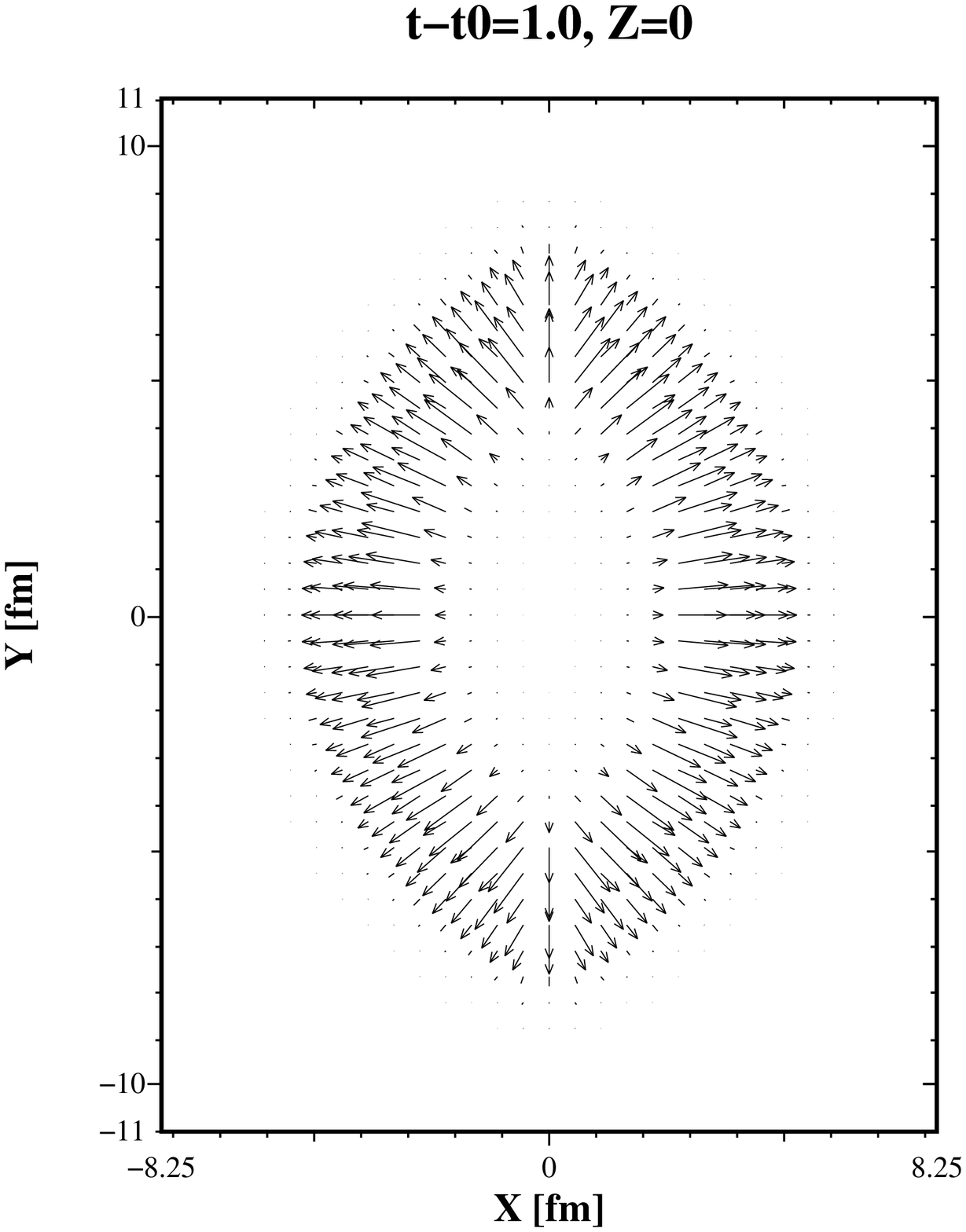}

\vspace{20pt}
\includegraphics[height=8cm]{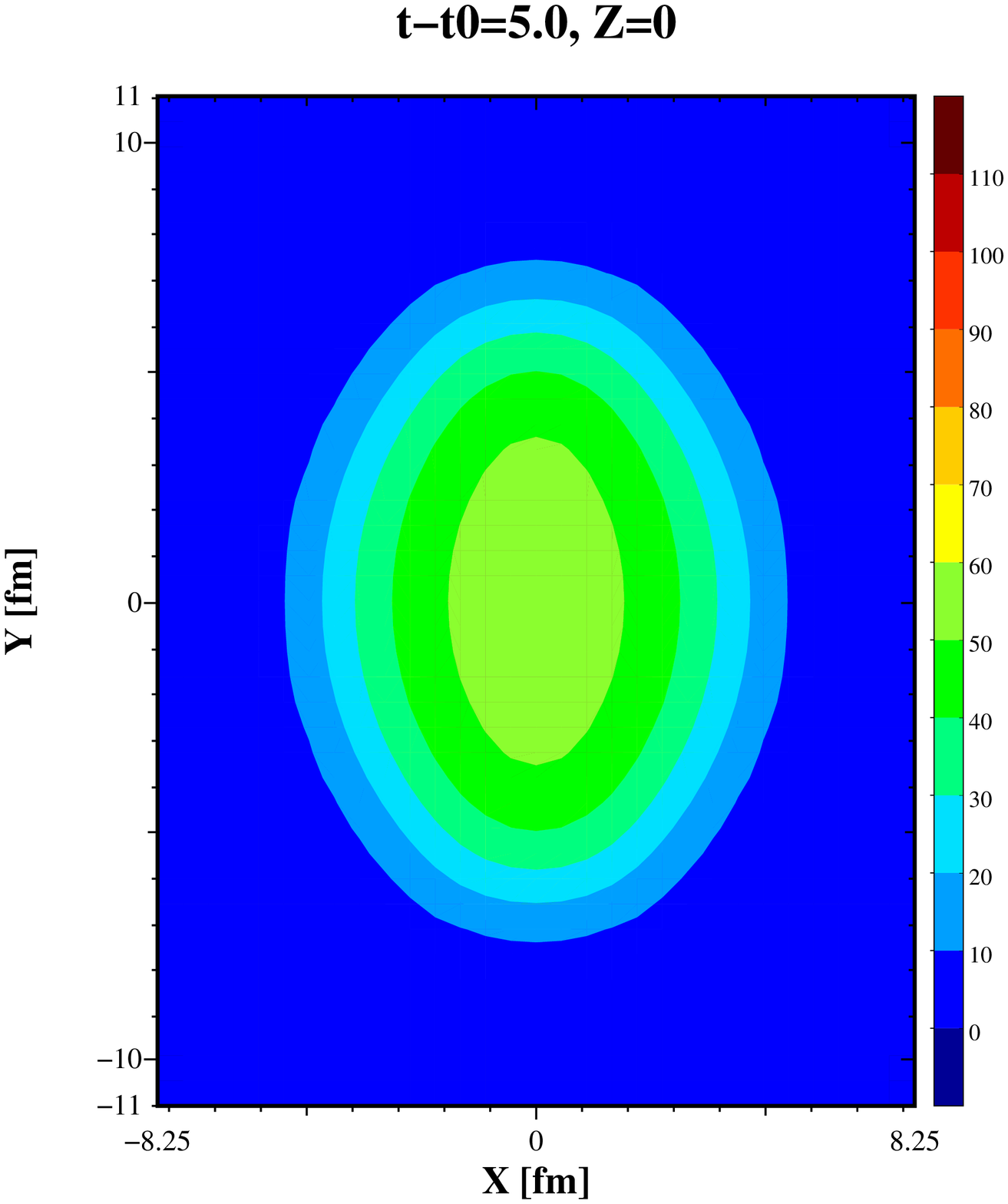}
\hspace{10pt}
\includegraphics[height=8cm]{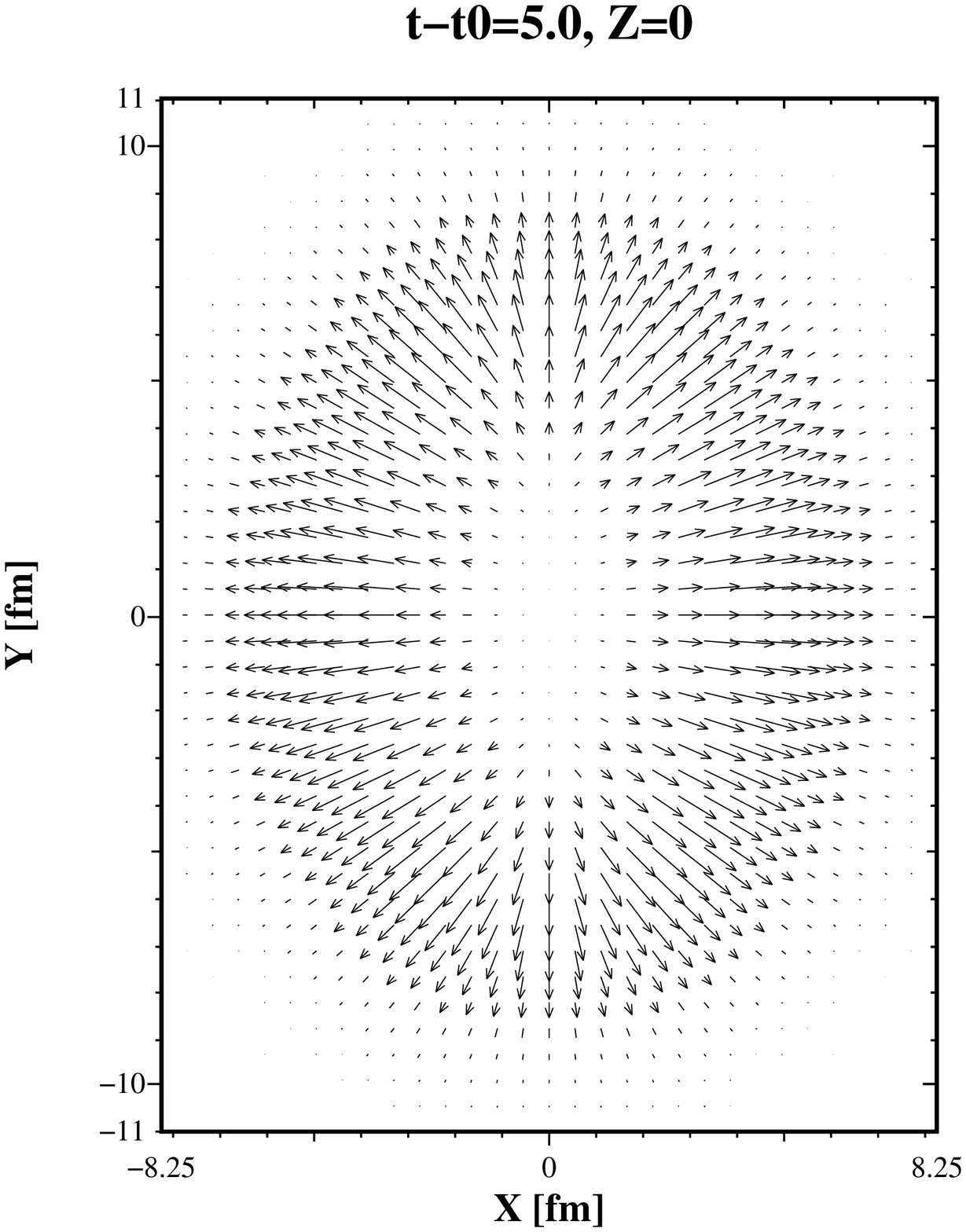}
\caption{Time evolution of pressure and baryon flow in the transverse plane.
Left: The pressure contours. Right: The baryon flow velocity vector
$(n_{\mathrm{\small{B}}}v_x, n_{\mathrm{\small{B}}}v_y)$.}
\label{fig:tempo}
\end{figure}
These parameters, however, should be chosen so as to reproduce the
experimental data for the (pseudo-)rapidity and the transverse
momentum distribution. 
To make our analysis more quantitative, we need this experimental data. We
would like the experimental group to analyze the \textit{centrality
dependence} of the hadron spectra, especially, the
(pseudo-)rapidity distribution. 
For this reason we wish to emphasize that our numerical results
presented below are only preliminary.


Figure \ref{fig:tempo} shows our numerical results for the temporal behavior
of the pressure (left column) and the baryonic flow (right
column) at $z=0$ in the non-central Pb+Pb collision with impact parameter $b=7$ fm at SPS energy.
Initially almost all matter in this plane is in the QGP phase and
there is no transverse flow anywhere by definition.
At $t = t_0 + 0.5$ fm we see the shell structure corresponding to the
mixed phase with the same pressure $\sim$ 70 MeV/fm$^3$, and the initial pressure gradient gives the baryons transverse flow.
The QGP phase disappears at $t = t_0 + 1.0$ fm and after that the mixed phase
occupies the central region.
There is still no transverse flow near the origin due to the absence of
a pressure gradient.
At about $t = t_0 + 5.0$ fm all the nuclear matter initially in the QGP phase
has gone through the phase transition and is in the hadronic phase.
We can see from these figures that the shape of the nuclear matter is
changing from almond (top figure on page 5) to round (bottom figure on page 6), and the elliptic flow reduces the
initial geometric deformation.

The numerical results of the hydrodynamical simulation give us the momentum
distribution through the Cooper-Frye formula \cite{CF} with freeze-out
temperature $T_{\mathrm{\small{f}}}=140$ MeV.
The elliptic flow parameter $v_2$, as a function of rapidity $y$,
is obtained from 
the momentum distribution
\begin{eqnarray}
v_2 (y) & = & \left<
\left(\frac{p_x}{p_t}\right)^2-\left(\frac{p_y}{p_t}\right)^2 \right>
\nonumber\\
    & = & \frac{\int_0^{2 \pi} d\phi \cos(2\phi) \int_{p_-}^{p_+} p_t dp_t
E\frac{d^3 N}{dp^3}}{\int_0^{2 \pi} d\phi \int_{p_-}^{p_+} p_t dp_t
E\frac{d^3 N}{dp^3}}.
\end{eqnarray}

Before calculating $v_2$ in non-central collisions with impact parameter
$b=7$ fm, we checked the numerical error in our hydrodynamic model in central
collisions. Since there is no special direction in the transverse plane for
head-on collisions, ideally the elliptic flow vanishes in the infinite particle
limit. Performing the numerical simulation with $b=0$ fm, we obtain the
value of $v_2$ as less than 10$^{-1}$ percent, therefore we can safely
neglect the numerical error. Note that the numerical error in the energy and
baryon density conservation of the fluid is less than one percent in our analysis.

\begin{figure}[htbp]
\begin{center}
\includegraphics[width=10cm]{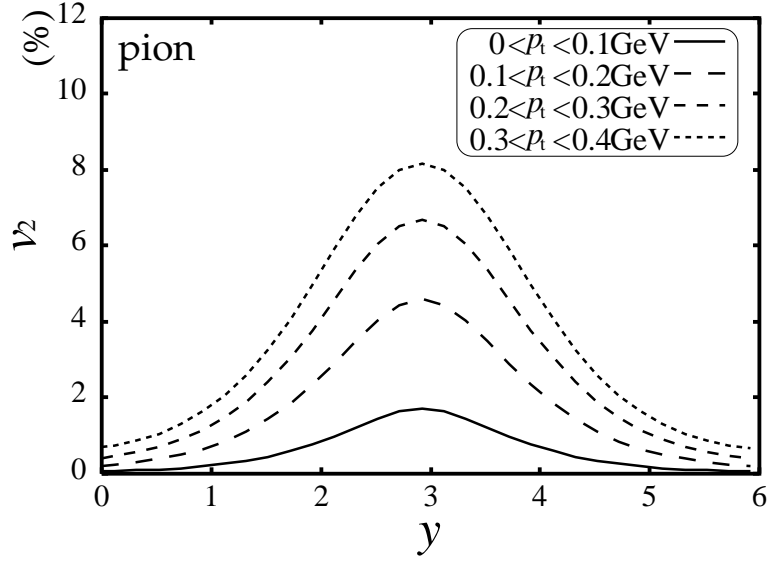}
\caption{Rapidity dependence of elliptic flow for pion. Four curves
correspond to the different transverse momentum regions. The midrapidity is
2.92.}
\label{fig:v2pi}
\end{center}
\end{figure}

Figure \ref{fig:v2pi} shows our results for the rapidity dependence of
elliptic flow for pions in different transverse momentum regions.
These results show that elliptic flow rises with transverse momentum
$p_t$ \cite{DANI} and has a peak at midrapidity.
\begin{figure}[htbp]
\begin{center}
\includegraphics[width=10cm]{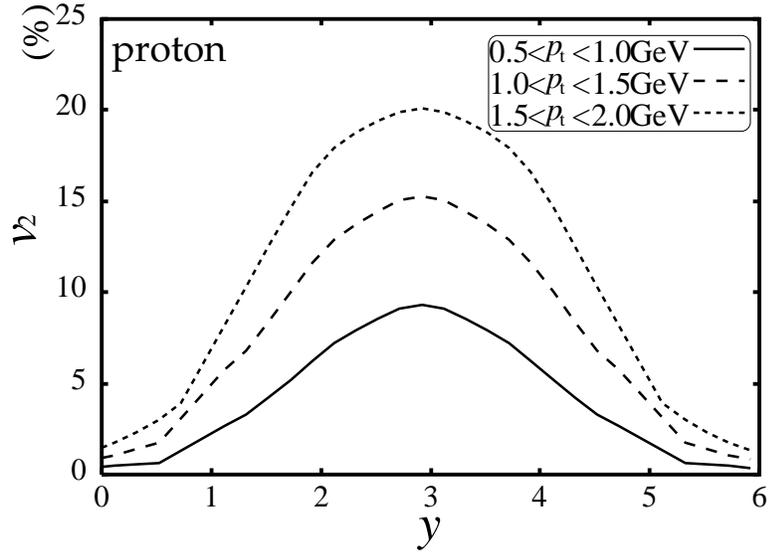}
\caption{Rapidity dependence of elliptic flow for proton. Three curves
correspond to the different transverse momentum regions. Note that the
integral region of transverse momentum is larger than for pions.}
\label{fig:v2pro}
\end{center}
\end{figure}
This seems to be in contrast with the experimental data obtained by the NA49
Collaboration \cite{NA49}. Their data appears to be slightly peaked at medium-high
rapidity.

Our results for $v_2$ for protons are shown in Fig.~\ref{fig:v2pro}.
We see the same behavior as for the pion case.
We obtain a larger $v_2$ for protons than for pions because we
are integrating over a larger transverse momentum region.
Since the initial parameters in our hydrodynamic model have been chosen by
hand, we would like readers to \textit{not} take these results
\textit{quantitatively}.

In summary, we reported our preliminary analysis of elliptic flow in
non-central heavy-ion collisions using the hydrodynamic model.
We numerically simulated the hydrodynamic model without assuming cylindrical
symmetry or Bjorken's boost-invariant solution, using the extended
version of the Piecewise Parabolic Method which is known as a robust scheme for
the non-relativistic gas equation with a shock wave. We presented the temporal
behavior of high temperature and high density nuclear matter produced in Pb+Pb
collisions with $b=7$ fm at SPS energy. Our preliminary results showed that
the elliptic flow parameter $v_2$ has a peak at midrapidity for both pions and
protons and increases with transverse momentum. Since there are some
ambiguities in the initial parameters of our hydrodynamical model, we should
fix these parameter using experimental data for the rapidity distribution in
non-central collisions. If we regard the hydrodynamical model as a
predictive one, we can choose initial parameters using results from a
parton cascade model, such as VNI \cite{GEIGER}. The study of
these issues is a future work.

The author is much indebted to Prof. I.~Ohba, Prof. H.~Nakazato, Dr. Y.
Yamanaka and Prof. S.~Muroya for their helpful comments, and to Dr.~H.~Nakamura, Dr.~C.~Nonaka and Dr.~S.~Nishimura for many interesting discussions. The numerical calculations were performed on workstations of the Waseda Univ.~high-energy physics group.

\end{document}